# Ultralow-current-density and bias-field-free spin-transfer nano-oscillator

Zhongming Zeng[1, 2], Giovanni Finocchio[3], Baoshun Zhang[1], Pedram Khalili Amiri[4], Jordan A. Katine[5], Ilya N. Krivorotov[6], Yiming Huai[7], Juergen Langer[8], Bruno Azzerboni[3], Kang L. Wang[4], and Hongwen Jiang[2]

[1]Suzhou Institute of Nano-tech and Nano-bionics, Chinese Academy of Sciences, Ruoshui Road 398, Suzhou 215123, P. R. China

[2]Department of Physics and Astronomy, University of California, Los Angeles, California 90095, United States

[3]Department of Electronic Engineering, Industrial Chemistry and Engineering, University of Messina, Messina 98166, Italy

[4]Department of Electrical Engineering, University of California, Los Angeles, California 90095, United States

[5]Hitachi Global Storage Technologies, San Jose, California 95135, United States

[6]Department of Physics and Astronomy, University of California, Irvine, California 92697, United States

[7]Avalanche Technology, Fremont, California 94538, United States

[8]Singulus Technologies, Kahl am Main 63796, Germany

**The spin-transfer nano-oscillator (STNO) offers the possibility of using the transfer of spin angular momentum via spin-polarized currents to generate microwave signals. However, at present STNO microwave emission mainly relies on both large drive currents and external magnetic fields. These issues hinder the implementation of STNOs for practical applications in terms of power dissipation and size. Here, we report microwave measurements on STNOs built with MgO-based magnetic tunnel junctions having a planar polarizer and a perpendicular free layer, where microwave emission with large output power, excited at ultralow current densities, and in the absence of any bias magnetic fields is observed. The measured critical current density is over one order of magnitude smaller**

**than previously reported. These results suggest the possibility of improved integration of STNOs with complementary metal-oxide-semiconductor technology, and could represent a new route for the development of the next-generation of on-chip oscillators.**

Oscillators are basic components of many communication, navigation, and measurement systems. They provide a periodic output that can be used to generate electromagnetic energy for radiation, for high speed digital systems, clocking functions, and to enable frequency up and down conversion if used as a local oscillator in communication circuits, and for many other applications. There is a strong interest, driven by cost and performance, to develop improved microwave oscillators for on-chip integration, and the spin-transfer nano-oscillator (STNO) is a promising candidate for this due to its combination of characteristics such as frequency tunability, nanoscale size, broad range of working temperature, and relatively easy integration with complementary metal-oxide-semiconductor (CMOS) technology[1-7]. The STNO is not only one of the more complex non-linear dynamical systems (of great interest from fundamental point of view), but it is also important being one of the smallest auto-oscillators known in nature (hence of great interest from technological point of view). The remaining challenges that need to be addressed are the simultaneous optimization of a variety of properties, including high frequency operation, frequency tunability, narrow spectral linewidth, large output power, and operation in both absence of external magnetic fields and with low applied current densities.

Here, we demonstrate that the interfacial perpendicular anisotropy (IPA) between the ferromagnetic electrodes and the tunnel barrier of magnetic tunnel junctions (MTJs) employing the material combination of CoFeB/MgO is a key ingredient to improve the dynamic properties of STNOs[8]. In general, IPA can be used for the design of high efficiency spintronic devices. It permits the realization of MTJs for storage applications with high thermal stability and low switching current[8, 9] and STNOs with very large emitted power (*i.e.* > 0.95 µW delivered to a matched load) [10]. In previous studies, IPA was used to orient the magnetization of both the free and polarizer layers out of the film plane[8], or to reduce the out-of-plane demagnetizing field ($H_d$) while maintaining the orientation of both of the two magnetizations in the film plane[10]. In this work, we designed STNOs with an asymmetric MTJ consisting of an in-plane polarizer and a perpendicular free layer achieved by controlling the trade-off between IPA and $H_d$. This orthogonal magnetic configuration is different from the one in our previous study[10], and is more efficient in order to excite large angle free layer precession (large output power) in absence of a bias magnetic field. The main objective of our experiment is to solve the key technological issues related to the integration of STNOs with CMOS technology: elimination of the need for an external magnetic bias field and reduction of drive current densities. The former goal will allow for eliminating the current lines necessary to create the magnetic field for the STNO biasing, while the latter is necessary to reduce the size of the current driving transistors (small currents allow for smaller transistors, hence making the overall oscillator smaller. For high-current oscillators, the actual size of the oscillator would be determined by the transistor size, rather than by the MTJ) and to control power dissipation.

**Results**

**Spin-transfer torque oscillator.** The samples studied have a core magnetic stack consisting of a synthetic antiferromagnet (SAF) $Co_{70}Fe_{30}/Ru/Co_{40}Fe_{40}B_{20}$ layer and a $Co_{20}Fe_{60}B_{20}$ free layer (FL) separated by a MgO insulator as shown in Fig. 1a. We introduce a Cartesian coordinate system where the x-axis is the direction of the polarizer (+x), the y-axis and the z-axis are the hard in-plane and the out-of-plane axes respectively. The SAF layer is designed to have an in-plane easy axis[9, 10] serving as a polarizer, while the Fe-rich $Co_{20}Fe_{60}B_{20}$ FL with a thickness of $t = 1.60$ ~1.62 nm was chosen to achieve the proper IPA, which favors the out-of-plane (perpendicular) magnetic configuration in the free layer (see Methods). Electron-beam lithography and ion milling were used to define and etch the MTJs resulting in pillar-shaped devices with nominal dimensions of 150 nm × 70 nm. The samples are different from the 50-70 nm diameter point contact spin valves with out-of-plane magnetized CoNi free layers studied in refs. 11, 12. A d.c. bias current is injected into the sample through a bias *Tee* as shown in the measurement setup in our previous work[10], where we define the positive current $I$ as electrons flowing from the polarizer to the free layer. A time-varying voltage produced by the oscillations of the magnetization via the tunneling magnetoresistance (TMR) effect is recorded using a 9 kHz-26.5 GHz spectrum analyzer. The measurements were carried out at room temperature. We show detailed data from a single device with free layer thickness $t = 1.60$ nm, however similar results have been achieved for more than 5 devices for each free layer thickness.

**Magneto-resistance transport properties.** Figure 1b shows the resistance of a typical nanopillar device as a function of the external in-plane field ($H_{||}$) at a sub-critical bias current of $I = 10$ μA, revealing a TMR ratio of 79%. As $H_{||}$ increases from -700 Oe to +1100 Oe, the resistance increases gradually as the magnetization of the free layer changes gradually from parallel to antiparallel relative to the polarizer. One source of asymmetry of the field scan is the dipolar coupling between the polarizer and the free layer, which (at low bias currents and zero external field) induces the free layer magnetization to tilt at a small angle away from z-axis (Fig. 1a). The additional switching near $H_{||} = -700$ Oe and +1100 Oe corresponds to reorienting of the polarizer magnetization. The resistance curve scan as function of the out-of-plane field $H_\perp$ (inset of Fig. 1b) clearly indicates the perpendicular free layer.

**Spin-transfer torque dynamics.** The devices exhibit dynamical behavior in a range of external fields (see Supplementary Fig. S1 online), but in the rest of the paper we will focus on the data measured at zero bias magnetic field, being the most important result of this experiment. Fig. 1c shows the resistance versus applied current $I$, an intermediate resistance (IR) state is seen in the range of $I$ = - 0.34 to + 0.65 mA, while the high resistance state (AP) for $I$ < - 0.34 mA and the low resistance state (P) for $I$ > 0.65 mA are observed. The large-power microwave emission is only observed in the IR state and for negative currents, *i.e.* B1 regime. We focused on frequency-domain measurements to establish the existence and study the characteristics (oscillation frequency $f_0$, linewidth $\Delta f_0$ and output power) of these STNOs.

Fig. 1d displays typical microwave power density spectra obtained for different values of $I$ for sample 1 ($t$ = 1.60 nm). For all of the current values ($|I| \leq 0.3$ mA), a single peak (with Lorentzian line shape) is observed in the GHz-range that is characterized by the excitation of an out-of-plane oscillation mode as indicated by the micromagnetic simulations (see Methods). The oscillation frequency exhibits a red shift, consistent with previous studies in point contact geometries with in-plane polarizer and out of plane FL[11], with a slope of 1.75 GHz mA$^{-1}$ and a minimum linewidth of 28 MHz (inset Fig. 1d). The estimated threshold current $I_c$ is ~ -45 µA (see Supplementary Note 2 online)[13], which corresponds to a current density of $J_c$ = -5.4×10$^5$ A/cm$^2$, this value can be further reduced by controlling the thickness of the free layer $t$, *e.g.* at $t$ = 1.62 nm $J_c$ = -1.2 × 10$^5$ A/cm$^2$ ($I_c$ = -10 µA), which is to the best of our knowledge the smallest value measured to date. The observation of the ultralow $J_c$ value can be explained as follows. In the presence of the IPA the $J_c$ is approximately proportional to the effective demagnetizing field[14] ($H_{eff} \approx H_{k\perp} - H_d$, where $H_{k\perp} > H_d$ is the IPA in the FL, which increases with decreasing the FL thickness[8, 9]). In our case, the strong IPA in the FL significantly counteracts the effect of $H_d$, resulting in the ultralow $J_c$, and is responsible for the strong dependence of $J_c$ on the FL thickness. As can be observed, the effect of the FL thickness on IPA has a non-trivial influence on the dynamical properties of the STNOs (for example a larger thickness corresponds to a smaller critical current for $H_{k\perp} > H_d$) and a complete systematic study will be presented elsewhere. Fig. 2 displays a comparison between the oscillation power as a function of the current for the two thicknesses $t$ = 1.60 and 1.62 nm. A complete set of data for $t$ =1.62 is included in the Supplementary Note 3 online (the dynamical properties of the STNO for $t$ = 1.60 and 1.62 nm are summarized in the last two rows of Table 1). The maximum oscillation power is larger at $t$ = 1.60 nm with the best result achieved for a current $I$ = - 0.3 mA where the measured output power is 18 nW (> 60 nW delivered to a matched load, see Supplementary Note 4 online) with an oscillation frequency and a linewidth of 850 MHz and 73 MHz, respectively. In the absence of external fields, an additional property is the reduction of total power dissipation, at the best working condition for $t$ = 1.60 nm the dissipated power is of the order of 50-70 μW which is a very small value for oscillators at sub-micrometer size. Hence, these results open the possibility to design nanoscale oscillators with improved dynamic performance and lower power dissipation.

**Comparison with simulations.** We performed micromagnetic simulations considering the same experimental framework to identify the origin of the steady-state magnetization oscillation. The simulations have been performed by solving the Landau-Lifshitz-Gilbert-Slonczewski equation (See Methods: Micromagnetic simulations). As in the experimental data, the magnetization precession is observed at negative currents and it is related to the excitation of a mode with an out-of-plane oscillation axis, and a trajectory which is initially circular and expands (i.e. the output power increases) as the current increases (compare trajectories at -82 and -164 µA in Fig. 3 (left)). At large currents the oscillation axis moves towards the x-y plane (trajectory at -288 µA) and for values of $I < -0.34$ mA the AP-state is obtained as in the experimental data. The magnetization dynamics are characterized by the excitation of a spatially quasi-uniform mode (see also Supplementary Video online). Fig. 3 (right) shows two examples of snapshots at $I = -0.3$ mA, the arrows indicate the in-plane component of the magnetization while the colors are related to the $m_x$ (blue negative, red positive). At high currents ($I < -0.3$ mA), the simulations show that for some range of time the dynamics is switched off (see Supplementary Fig. S4 online). We believe that the presence of this behavior could be the origin of the non-uniform dynamics (experimental low frequency tail) measured in the high current regime (see Supplementary Fig. S3 online). We observed that this behavior is more evident in the range of thickness where the IPA is comparable with the out-of-plane demagnetizing field. The simulations also predict a decrease in the oscillation frequency as a function of $I$ at a rate of ~ 1.8 GHz mA$^{-1}$, this value is consistent with the experimentally observed rate of ~ 1.75 GHz mA$^{-1}$ (Fig. 4).

## Discussion

Microwave emission with large output power (in the presence of external magnetic fields and large drive current densities) has been already measured in STNOs with IPA[10, 15]. However, to the best of our knowledge this work is the first experimental demonstration of large oscillation power without bias field and with ultralow current density $J_c$ below $6\times10^5$ A/cm$^2$. For microwave generation in the absence of external magnetic fields, various other solutions have been recently proposed. For example, the idea based on a wave-like angular dependence of the spin torque[16], the incorporation of a perpendicular polarizer into spin-valve structures[17], magnetic vortex oscillations[6, 18] or a tilted free layer[19], but the resulting microwave power in those STNOs is smaller (<1 nW) and the excitation current is higher (> $10^6$ A/cm$^2$) [11, 16-18]. When compared with the best results obtained from previous field-free STNOs, the data in this work show microwave emission with output power at least one order of magnitude larger and critical current densities one order of magnitude smaller. Table 1 compares the dynamical properties of STNO solution achieved without magnetic field. Importantly, the oscillation frequency is of the same order as that obtained from other types of STNOs. Finally, the STNOs in this work exhibit a tunability of ~ 1.75 GHz mA$^{-1}$ that is substantially larger than the ones measured in spin-transfer driven vortex oscillations (0.03 GHz mA$^{-1}$ in Ref. 6 and 0.08 GHz mA$^{-1}$ in Ref. 18). This, however, comes at the cost of the larger linewidth compared to those obtained in magnetic vortex self-oscillations[6, 18]. The origin of the linewidth is related primarily to thermal fluctuations, and to the coupling between the oscillator phase and power as in other STNOs[13]. One possible remedy is the use of these STNOs as a basis of an array of phase locked STNOs, in that scenario it is expected that a significant decrease in the linewidth and an increase in the output power to over the tens of μW[20, 21] could be achieved. A different solution would be the application of a low frequency current modulation as demonstrated in Ref. 22.

The possibilities opened by these results will eliminate some of the key issues related to on-chip integration of STNOs with CMOS technology. This direct integration and the reduced power consumption may potentially open applications in portable electronic devices and wireless modules such as embedded communications and power efficient local clock signal generation in digital systems. Our findings also provide a key ingredient in the development of ultralow–critical-current and zero-field spin-wave sources in magnonic logic devices[23].

**Methods**

**Sample preparation**. The magnetic stacks were deposited by sputtering in a Singulus TIMARIS PVD system and contain a layer structure: PtMn (15) /Co$_{70}$Fe$_{30}$ (2.3) /Ru (0.85) /Co$_{40}$Fe$_{40}$B$_{20}$ (2.4) /MgO (0.8) /Co$_{20}$Fe$_{60}$B$_{20}$ (1.5 ~ 2.0) (the thicknesses are in nm). Details of the growth and fabrication process are similar to that in our previous work[9]. Note that the IPA gradually increases as the Co$_{20}$Fe$_{60}$B$_{20}$ FL thickness (*t*) decreases[8, 9]. The direction of the FL magnetic moments is mainly determined by the competition between the IPA ($H_{k\perp}$) and the $H_d$. For a thicker FL, e.g. *t* =1.70 nm[10], it has been verified that the magnetization is in-plane and it requires an external magnetic field to produce microwave signals. By contrast, the thinner *t* < 1.60 nm results in a larger $H_{k\perp}$ and consequently a larger critical current density. We find that, the microwave oscillations with no external field exist in a thickness range from 1.55 to 1.65 nm, and that in particular, for the FL thickness between 1.60 and 1.62 nm, better microwave performance in terms of power and drive current density is achieved.

**Micromagnetic simulation**. We numerically solve the Landau-Lifshitz-Gilbert-Slonczewski equation which also includes the field-like torque term $T_{OP}$ [24, 25]. $T_{OP}$ is considered dependent on the square of the bias voltage[26] up to a maximum value of 25% of the in-plane torque computed for a current density $|J|$ = 6.0×10$^6$A/cm$^2$. The total torque, including also the in-plane component $T_{IP}$ is:

$$T_{IP} + T_{OP} = \frac{g|\mu_B|J(\mathbf{m},\mathbf{m_p})}{|e|\gamma_0 M_s^2 t} g_T(\mathbf{m},\mathbf{m_p})[\mathbf{m}\times(\mathbf{m}\times\mathbf{m_p}) - q(V)(\mathbf{m}\times\mathbf{m_p})], \qquad (1)$$

where *g* is the gyromagnetic splitting factor, $\gamma_0$ is the gyromagnetic ratio, $\mu_B$ is the Bohr magneton, *q(V)* is the voltage dependent parameter for the perpendicular torque, $J(\mathbf{m},\mathbf{m_p})$ is the spatial dependent current density, *V* is the voltage (computed from Fig. 1c in the main text), *t* is the thickness of the free layer, and *e* is the electron charge[27]. The effective field takes into account the standard micromagnetic contributions (exchange, self-magnetostatic) and the magnetic coupling between free and polarizer layer as well as the Oersted field due to *I*. The presence of the IPA has been modeled as an additional contribution to the effective field. The parameters used for the CoFeB have been: saturation magnetization $M_s$ = 1.1×10$^6$ A/m, IPA constant $k_u$ = 7.4×10$^5$ J/m$^3$, exchange constant $A$ = 2.0×10$^{-11}$ J/m, damping parameter $α$ = 0.01. The polarization function $g_T(\mathbf{m},\mathbf{m_p}) = 2\eta_T(1+\eta_T^2\,\mathbf{m}\cdot\mathbf{m_p})^{-1}$ has been computed by Slonczewski[28,29], being not dependent on the bias voltage, where **m** and **m$_p$** are the normalized magnetizations of the free and polarizer layer, respectively. To fit the critical current density, we use for the spin-polarization $\eta_T$ the value 0.57[30]. Micromagnetic simulations provide useful insight into the nature of the excited mode in the STNO with IPA. They indicate the excitation of an out-of plane mode characterized by a quasi-uniform spatial distribution of the magnetization.

**Acknowledgements**

This work was supported by 100 Talents Programme of The Chinese Academy of Sciences, the National Science Foundation of China (11274343), the DARPA STT-RAM and Non-Volatile Logic programs, and the Nanoelectronics Research Initiative (NRI) through the Western Institute of Nanoelectronics (WIN). This work was also supported by a Spanish Project under Contract No. MAT2011-28532-C03-01.

## Author contributions

Z. M. Z., P. K. A., I. N. K. and H. W. J. designed the experiment, J. L. deposited the films, and J. A. K. fabricated the devices. Z. M. Z. collected and analyzed the data, G. F. performed simulations and analysis. Z. M. Z. and G. F. prepared the manuscript. All authors discussed the results, contributed to the data analysis, and commented on the final manuscript.

## Additional information

**Supplementary information** accompanies this paper at http://www.nature.com/Scientificreports

**Competing financial interests**: The authors declare no competing financial interests.

**License**: This work is licensed under a Creative Commons Attribution-NonCommercial-NoDerivs 3.0 Unported License. To view a copy of this license, visit

http://creativecommons.org/licenses/by-nc-nd/3.0/

**How to cite this article:** Z. M. Zeng *et al.* Ultralow-current-density and bias-field-free spin-transfer nano-oscillator. *Sci. Rep.* **3**, 1426 (2013).

## Figure captions

**Figure 1. Sample structure and properties.** (a) Schematic of the sample layer structure consisting of an in-plane magnetized fixed (polarizer) layer and an out-of-plane magnetized free layer. (b) Resistance as a function of in-plane magnetic field ($H_{\parallel}$) and perpendicular magnetic field ($H_{\perp}$) for sample 1 ($t = 1.60$ nm), inset in (b) is the resistance as a function of $H_{\perp}$, the black (red) arrow denotes the magnetization direction of the reference (free) layer. (c) Resistance-Current curve at zero applied magnetic field, AP (P) denotes the antiparallel (parallel)

configurations between the free and fixed layers. (d) Microwave spectra as a function of d.c. current bias $I$ at zero applied magnetic field, the curves are offset by approximately 20 nW GHz$^{-1}$ along the vertical axis for clarity. Inset: full width at half maximum (FWHM, or linewidth) (triangles) and $f_0$ (circles) of the STNO sample 1 as a function of $I$.

**Figure 2**. **Power of STNO samples**. Integrated power of fundamental signals at zero applied magnetic field as a function of $I$. Olive circles are for 1.60 nm sample and blue squares for 1.62 nm sample.

**Figure 3**. **Micromagnetic simulations for sample 1.** Left. Trajectories of the average magnetization vector on the unit sphere as computed from micromagnetic simulations ($I$ = -82, -164 and -288 µA). Right: example of two snapshots of the spatial distribution of the magnetization indicating the uniform dynamics (the color means the x-component of the magnetization blue negative, red positive).

**Figure 4**. **Dependence of microwave frequencies on current for sample 1**. The blue circles show experimental data as a function of current bias at zero applied magnetic field. The black squares show the results from micromagnetic simulations.

**Table 1**

Comparison among the performance parameters (external field, critical current density, oscillation power, oscillation frequency, and the minimum linewidth) of different STNO solutions and the ones reported in this work at zero field and for two thicknesses $t$ (1.60 nm and 1.62 nm). Here $P_{mes}$ is the measured maximum power integrated from the oscillation peak, $P_{max}$ is the maximum power delivered to a matched load, and the hyphen (-) indicates cases where no data are available.

| STNOs | $H$ (Oe) | $J_c$ (A/cm$^2$) | $P_{mes}$ (nW) | $P_{max}$ (nW) | $f_0$ (GHz) | $\Delta f_{0min}$ (MHz) |
|---|---|---|---|---|---|---|
| Ref. 11 | 2500 | $> 3.0\times10^7$ | $< 1.0$ | - | 6 ~ 10 | 6 |
| Ref. 16 | $\geq 2$ | $> 5.3\times10^7$ | $< 0.003$ | - | 2 ~ 3 | - |
| Ref. 17 | $\geq 300$ | $> 9 \times10^6$ | $< 0.1$ | - | 2 ~ 4 | - |
| Ref. 18 | $> 4500$ | $> 4.4\times10^6$ | 5.0 | - | ~ 0.78 | 1.1 |
| This work $t = 1.60$ nm | 0 | $< 5.4\times10^5$ | 18 | 63 | 0.6 ~ 1.5 | 28 |
| This work $t = 1.62$ nm | 0 | $< 1.2\times10^5$ | 11 | 36 | 0.6 ~ 1.0 | 33 |

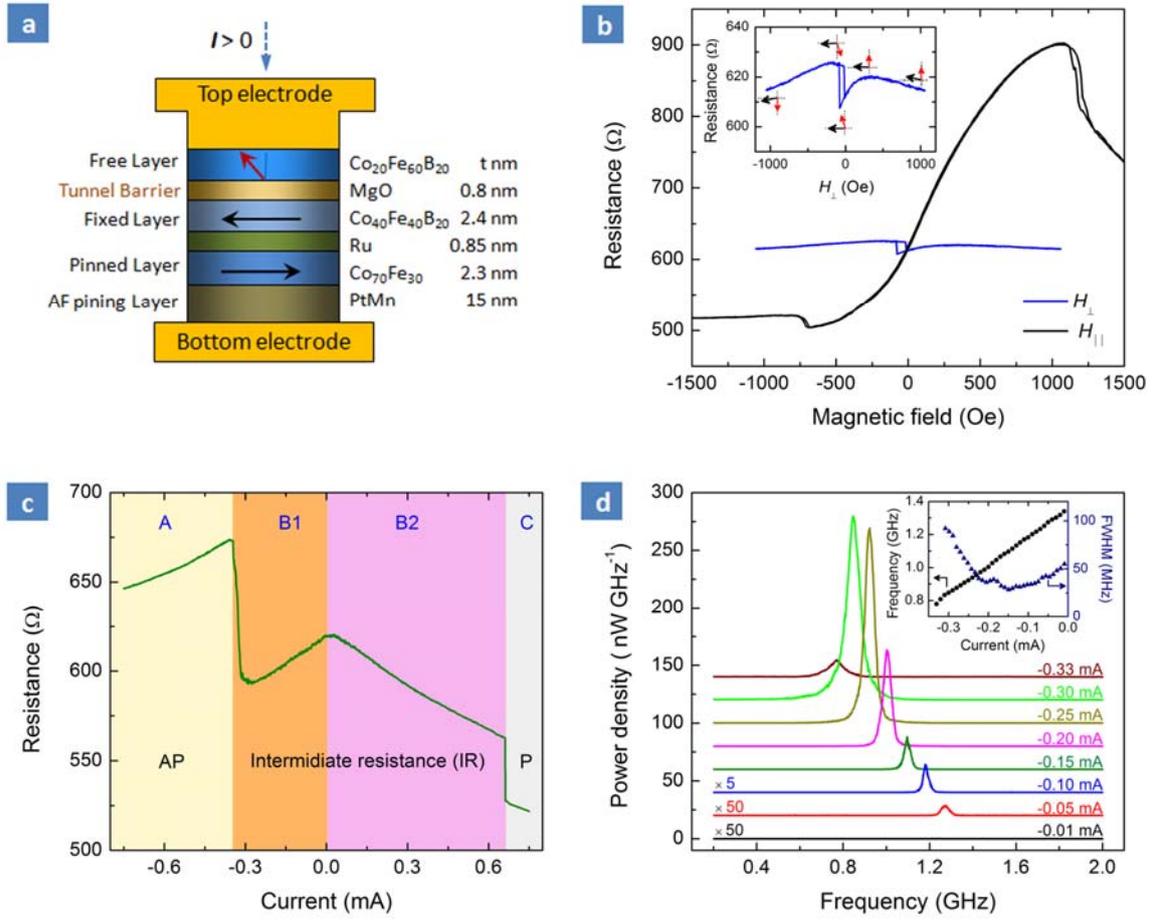

Figure 1

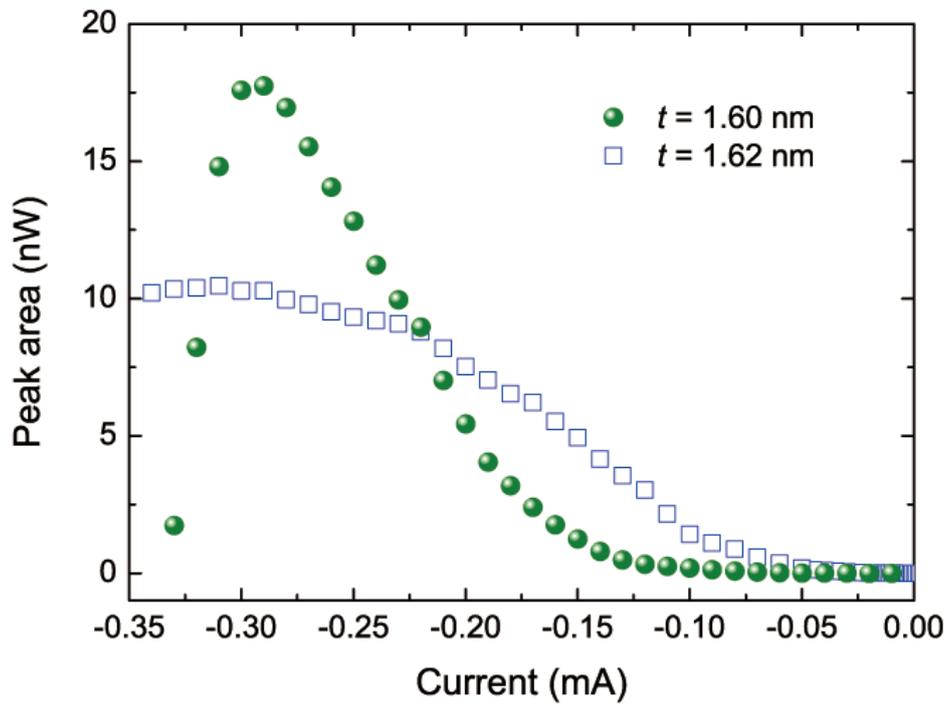

Figure 2

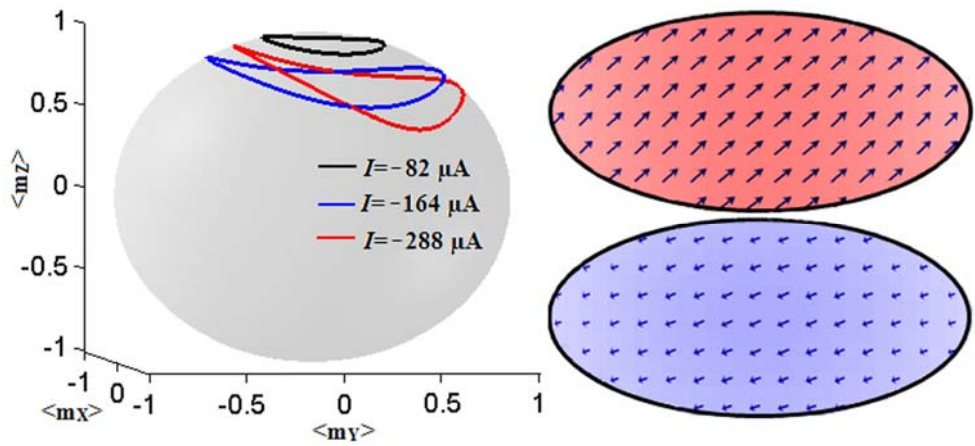

Figure 3

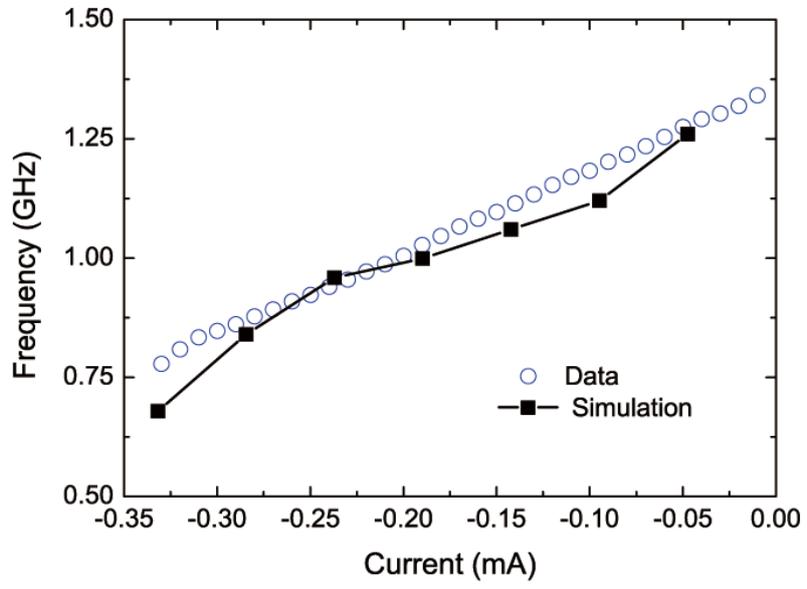

Figure 4